\documentclass{aastex631}
\usepackage{amsmath}
\usepackage{comment}
\usepackage{comment}
\usepackage{amsbsy}

\newcommand{\nflares}{{216}}
\newcommand{\ncflares}{{480}}
\usepackage{booktabs}

\shorttitle{}
\shortauthors{Kazachenko et al.}
\usepackage{stackengine}

\begin{document}
\title{{A Database of Magnetic and Thermodynamic Properties of Confined And Eruptive Solar Flares} }

\author[0000-0001-8975-7605]{Maria~D.~Kazachenko}
\affiliation{Dept. of Astrophysical and Planetary Sciences, University of Colorado Boulder,  \\
 2000 Colorado Ave, Boulder, CO 80305, USA}
\affiliation{National Solar Observatory,\\
 3665 Discovery Drive, Boulder, CO 80303, USA}




\email{maria.kazachenko@colorado.edu}
\begin{abstract}

Solar flares sometimes lead to coronal mass ejections that directly affect the Earth's environment. {However, a large} fraction of flares, { including} on solar-type stars, are confined flares. What are the differences in physical properties between confined and eruptive flares?  For the first time, we quantify thermodynamic and magnetic properties of hundreds of confined and eruptive flares of GOES class C5.0 and above,  $480$ flares total. We first analyze large flares of GOES class M1.0 and above observed by the Solar Dynamics Observatory (SDO): {$216$ flares total, including $103$ eruptive and $113$} confined flares, from 2010 until 2016 April,
we then look at the entire dataset above C5.0 of $480$ flares. We compare GOES X-ray thermodynamic flare properties, including peak temperature and emission measure, and active-region and flare-ribbon magnetic field properties, including reconnected magnetic flux and peak reconnection rate.
We find that for {fixed peak X-ray flux}, confined and eruptive flares have similar reconnection fluxes; however, {for fixed peak X-ray flux} confined flares have {on average} larger peak magnetic reconnection rates, are more compact, and occur in larger active regions than eruptive flares. 
These findings suggest that confined flares are caused by reconnection {between more compact, stronger, lower lying magnetic-fields} in larger active regions that reorganizes smaller fraction { of these regions' fields}. This reconnection proceeds at faster rates and ends earlier, { potentially leading} to more efficient flare particle acceleration in confined flares. 

  \end{abstract}
\keywords{Sun: flares -- Sun: magnetic fields -- Sun: ARs}
%
%

\section{Introduction}\label{intro}
Flares are classified into eruptive and confined types according to their association with coronal mass ejections (CMEs, \citealt{Moore2001,WangZhang2007,Thalmann2015,Temmer2021,Afanasyev2023}). In an eruptive flare, plasma is ejected into interplanetary space and is later observed as a CME in white-light coronagraph images, although its visibility might greatly depend on flare intensity \citep{Yashiro2006}. In a non-eruptive (or confined or compact) flare, plasma falls back to the Sun and there is no  CME.  On the Sun both types of flares are frequent: for example, from 2010 to 2019 only half of large solar flares of GOES class M1.0 and above flares were eruptive and led to CMEs; another half were confined flares \citep{Li2020}.  On solar-type cool stars, {{ up to} now there have been} only $40$ CME detections \citep{Moschou2019,Veronig2021,Namekata2022}. 

\begin{figure*}[tb!]
    \begin{center}
     \centerline{
    \includegraphics[width=1.0\textwidth]{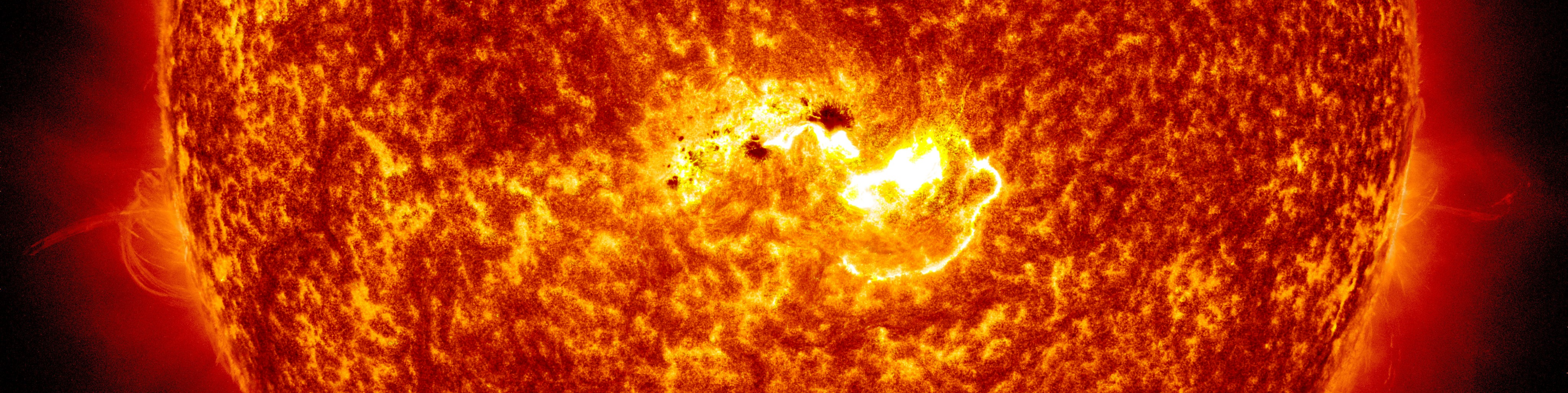} 
    }
     \caption{X1.2-class eruptive solar flare observed by the Solar Dynamics Observatory in 1600\AA{} and 304\AA{} wavelengths. Bright ribbons mark the footpoints of reconnected fields lines thaft provide indirect measurement of the amount of reconnecting magnetic flux in the flare and { the reconnection rate}.}
 \label{figure:sdo} 
\end{center}
\end{figure*}

While comparison of confined and eruptive flares on solar-like stars is difficult due to lack of spatial resolution, on the Sun we can measure both magnetic and thermodynamics properties of flares in high detail (see Figure~\ref{figure:sdo}).  With the launch of the Solar Dynamics Observatory in 2010, the number of studies comparing properties of eruptive and confined flares surged. For example, \citet{Cheng2011} analyzed $9$ M- and X-class flares all from the same active region (AR) to understand magnetic and thermodynamic properties of $6$ confined and $3$ eruptive events. {\citet{Harra2016} analyzed properties of $9$ confined and $33$ eruptive flares, $42$ X-class flares total.} \citealt{Hinterreiter2018} and \citet{Chernitz2018} analyzed $50$ flares, $19$ eruptive and $31$ confined events, C-class and above. \citet{Toriumi2017} surveyed $51$ flares, $32$ eruptive and $19$ confined events, M5.0-class and above. Finally, \citet{Li2020} analyzed $322$ flares, $170$ eruptive and $152$ confined events, M1.0-class and above, followed by the largest-to-date study, by \citet{Li2021}, of $719$ flares, $251$ eruptive and $468$ confined events, C5.0-class and above.

{Two physical} concepts are typically used to describe whether a flare would be eruptive or confined. The first factor describes the degree of AR non-potentiality, i.e. magnetic free energy, twist, helicity, etc. The second factor describes the constraining effect of the strapping or overlying field: its strength and decay rate with height.  { The} torus instability occurs once the decay index of $n=-\frac{\partial \mathrm{ln} {B}_\mathrm{p}}{\partial \mathrm{ln} z}$ for poloidal field ${B}_\mathrm{p}$ reaches a critical eruptive value of $n_\mathrm{cr}$ at a critical height $h_\mathrm{cr}$. 
{In this case the poloidal component refers to the component of the strapping field perpendicular to the axis of the toroidal flux rope.}
 {For critical values, a range} within $0.5$ to $2$ is typically used, depending on the magnetic configuration  (see e.g., \citealt{Torok2005,Kliem2006,WangZhang2007,Myers2015,Sun2015,Sun2022,Hassanin2016,Hassanin2022} and references therein). In particular, a value of $n_c=1.5$ for a thin, axisymmetric torus \citep{Bateman1978} is widely used in observational studies. As larger { total unsigned} AR magnetic flux leads to increased critical decay-index height $h_\mathrm{cr}$, ARs with large magnetic flux have confined eruptions due to strong magnetic cages (e.g., \citealt{Amari2018,Li2021}). 
 
Below we {review some magnetic field properties quantified} in confined and eruptive events from observations. {For more details on recent statistical studies of flare magnetic properties, we recommend reading our recent review paper \citep{Kazachenko2022r}}.

The main magnetic-field properties {typically used} to distinguish confined from eruptive events {include} { total unsigned} magnetic fluxes of flare-hosting  ARs, $\Phi_\mathrm{AR}$ \citep{Li2021}, fractions of the active-region magnetic flux or area {swept by flare ribbons, $\Phi_\mathrm{rbn}$ and $S_\mathrm{rbn}$, relative to} the total AR flux or area, $R_\Phi=\Phi_\mathrm{rbn}/\Phi_\mathrm{AR}$ and $R_S=S_\mathrm{rbn}/S_\mathrm{AR}$, respectively \citep{Thalmann2015,Masson2017,Chernitz2018,Toriumi2017,Li2020, Kazachenko2022}, and the ratio of the twist {near} the flaring polarity inversion line (PIL) to the AR magnetic flux, $\alpha_\mathrm{PIL}/\Phi_\mathrm{AR}$ (relative non-potentiality, \citealt{Li2022}). 
{For a given flare class, eruptive flares involve} larger fraction of AR magnetic flux and area, 
tend to occur in smaller ARs and have a larger ratio of the twist {near the PIL} to the AR flux.
 {Using a sample of $43$ eruptive and $63$ confined flares, \citet{Li2022} found that the relative non-potentiality parameter of $\alpha_\mathrm{PIL}/\Phi_\mathrm{AR}>2.2\times 10^{-24}Mm^{-1}Mx^{-1}$  {distinguishes} about $93\%$ of eruptive events}.
On the other hand, confined flares have smaller fractions of AR magnetic flux and area, 
tend to occur in larger ARs and have a smaller ratio of the twist {near the PIL} to the AR flux. Again, in the work by \citet{Li2022} the above relative non-potentiality of $\alpha_\mathrm{PIL}/\Phi_\mathrm{AR}<2.2\times10^{-24}Mm^{-1}Mx^{-1}$ {distinguishes} about $83\%$ of confined events. 
{In addition, from analysis of} photospheric vector magnetic fields, \citet{Liu2017,Vemareddy2019,Avallone2020}, and \citet{Kazachenko2022} found that confined flares tend to occur in ARs that are more current-neutralized, and have a smaller amount of magnetic shear at {the} PIL.  As shear and net current are {proportional to each other through Ampere's law}, smaller shear and net current in confined flares both support the same physical concept of smaller magnetic-field non-potentiality at the PIL \citep{Kazachenko2022}.
%
%
To summarize, {existing analyses of magnetic field observations} fit within the above scenario that smaller non-potentiality and stronger overlying fields in confined flares constrain plasma from escaping. In terms of magnetic topology of pre-flare magnetic field, both confined and eruptive flares could originate from a sheared arcade or a flux rope \citep{Masson2017} with one or several confined flares preceding an eruption \citep{Kliem2021,Hassanin2022}.

Numerous studies found that proximity of the flaring {part} of the AR to the AR boundary and the open field affect the flare eruptivity.  For example, \citet{Chernitz2018} found  that for a given X-ray flux, confined events tend to have larger mean magnetic-flux density, implying that they tend to occur closer to the center of ARs where fields are stronger.  \citet{Cheng2011} came to a {similar conclusion,} finding that 
{confined flares occur closer to AR centers than eruptive flares, in agreement with}
 \citet{WangZhang2007}.  
{Using non-linear force-free field extrapolation, they also found that in the low corona ($\approx10$ Mm), eruptive flares have a higher decay index of the transverse magnetic field than confined flares}.
  In addition, the strength of the transverse magnetic field over the eruptive flare sites is weaker than it is over the confined ones in agreement with e.g., \citet{Cui2018,Vasantharaju2018}. 
Finally, \citet{Derosa2018} analyzed the PFSS global magnetic fields. { Using a sample of $50$ eruptive and $6$ confined flares, they found that confined events have slightly less access to open flux than eruptive events. However this difference is not large: the rate at which X-class flares are eruptive is  $0.97$ (30 out of 31 flares) and $0.8$ ($20/25$) for events with and without access to open flux, respectively.} This result agrees with \citet{Baumgartner2018}.  In addition, {an active region} could become more prone to host  {a confined flare} when the orientation between {its} local and overlying magnetic flux systems becomes less antiparallel \citep{Zuccarello2017}.

%
Among {\it other properties} differing in confined and eruptive flares, average ribbon-separation distances and ribbon-peak separation speeds tend to be smaller for confined flares \citep{Kurokawa1989,Su2007,Veronig2015,Thalmann2015,Hinterreiter2018}.  
{ One explanation for this could be strong overlying fields in confined events which prevent reconnecting current sheet from moving upwards.} As ribbon separation speed is a proxy for a localized reconnection rate, results by \citet{Hinterreiter2018,Thalmann2015} imply that eruptive flares {might} have higher local reconnection rate, {however the authors of these papers do not explicitly state { this}}.  
  Note {also}, that while \citet{Hinterreiter2018} analyzed a large number of flares of different flare classes ($50$ flares, $31$ confined and $19$ eruptive), ranging from B- to X-class, their {distributions of GOES peak X-ray fluxes of confined and eruptive flare samples were  different,} with the confined flare sample being significantly {smaller} than eruptive sample.  

While a large number of statistical studies compared confined and eruptive flares' magnetic properties, few have analyzed thermodynamic properties and energy partition of confined flares and even fewer have compared those with eruptive events.
\citet{Kahler2022} analyzed GOES peak temperatures in hundreds of flares, {of flare class M3.0 and above}, and found, that for a given peak X-ray flux,  confined flares have higher temperatures than eruptive {flares, confirming} earlier results of  \citet{Kay2003} study of $69$ flares. 
\citet{Cheng2011} analyzed $9$ M- and X-class flares from the same ARs: $6$ confined and $3$ eruptive events. They found that 
{confined flares have more impulsive soft X-ray time profiles, while eruptive flare have longer durations and extend { over} larger areas in EUV 195 \AA{} images.}
{\citet{Harra2016}  analyzed $42$ flares, X-class and above: $9$ confined and $33$ eruptive flares. For their sample they found no statistical differences in thermal and magnetic properties between confined and eruptive flares.} 
%
 \citet{Tan2021} analyzed radio and X-ray data from GOES, RHESSI, and Fermi/GBM from 10 mostly confined flares in AR 12192. They found that the magnetic field in confined flares tends to be stronger than that in  {$412$ eruptive flares studied} by \citealt{Nita2004}. They also found that confined flares are efficient particle accelerators, however the energies to which electrons are accelerated in confined flares are lower (and do not produce HXR above $300$ keV, indicating the lack of efficient acceleration of electrons to high energies), in agreement with \citet{Thalmann2015}.  
%
\citet{Cai2021} analyzed energy partition in $4$ confined flares. They found that the ratio of non-thermal energies to magnetic energies is significantly larger for confined than for eruptive flares, ranging within $E_\mathrm{nth}/E_\mathrm{mag}\approx0.7-0.76$, in agreement with case study by \citet{Thalmann2015}. This implies that larger fraction of free energy is converted into kinetic energy of flare-accelerated electrons in confined flares, while a larger fraction of free energy is converted into other forms of energy, {such as kinetic energy of particles, thermal energy, and CME kinetic energy, in} eruptive flares \citep{Emslie2005,Reeves2010,Emslie2012,Warmuth2016,Aschwanden2015,Aschwanden2017}.  {Given that CME kinetic energy is the largest fraction in the eruptive-flare energy partition, it is not surprising that eruptive flares have smaller energy fractions in accelerated particles, compared with confined flares.} 
Finally, \citet{Qiu2022} analyzed three eruptive (M2.0, M6.9 and M1.5) and three confined (C8.2, C8.7 and C9.8) flares. They found that in eruptive flares non-thermal HXR emission lags the EUV emission from flare ribbons, suggesting that eruptive flares have a gradual warm-up phase with lower non-thermal energy release efficiency. They also found that, on average, confined flares exhibit stronger magnetic shear and high-temperature component (up to $25$ MK for up to one minute) at the onset, which is not present in eruptive flares (in agreement with case study by \citealt{Veronig2015}). Note, however, that eruptive flares { in the study by \citet{Qiu2022}} had larger flare classes than confined events; ideally, flare samples of similar flare classes should be compared.  
To summarize, confined flares are efficient particle accelerators located lower in the corona in agreement with earlier studies (e.g., \citealt{Klein2010}), with only a small fraction of particles accelerated to extremely high energies.

While most statistical papers focused on magnetic properties of eruptive/confined flares and few analyzed their thermodynamic properties,
in this paper {\it our goal} is to fill this gap to create a comprehensive survey of both thermodynamic and magnetic properties of eruptive and confined flares on the Sun {for a large, balanced sample}. For { this,} we analyze properties of all flares {{ of GOES class C5.0}} and above observed by the SDO from 2010 to 2016 within $45^\circ$ from the {central meridian}. %
{To minimize errors in the reconnection fluxes we restrict our analysis to events with reconnection flux imbalance between positive and negative polarities $\Phi_\mathrm{ribbon,imb}<20\%$. }
 As a result we select {$480$ events total: $152$ eruptive and $328$ confined flares of GOES class C5.0 and above. Since  the number of confined and eruptive flares in this sample is highly unbalanced, we first analyze a more balanced subsample of large flares of GOES class M1.0 and above,}  {$216$} events total: {$103$} eruptive and {$113$} confined flares. {We then present an analysis for the whole dataset, coming to similar physical conclusions (see Appendix~\ref{app}).}
We analyze the following flare  properties: AR magnetic fluxes, magnetic-reconnection fluxes and their peak rates, flare thermodynamic properties (temperature, emission measure) and flare durations. 

This paper is organized as follows. In Section~\ref{methodology}, we describe {the four datasets} that we use to assemble our final dataset, the variables and the procedures to evaluate uncertainties. In Section~\ref{results} we discuss our results for $216$ flares of GOES class M1.0 and above.  {For comparison, in Appendix~\ref{app} we present results for a larger sample of $480$ flares that includes smaller flares above C5.0}.  Finally, in Section~\ref{conc}, we summarize our conclusions {for both flare samples}.

\section{Data and Methodology}\label{methodology}

Table~\ref{vars_tab} shows a list of physical properties and their data source that we use for each flare in our analysis. These properties include flare location, GOES peak X-ray flux, {GOES time-integrated flux (fluence)}, GOES start, peak, end times and durations; GOES peak temperature and emission measure; AR area and unsigned magnetic flux; unsigned reconnection flux, peak reconnection rate, ribbon area and eruptivity (confined or eruptive) of each flare.

{For initial flare detections and estimates of flare} reconnection flux, ribbon area and their uncertainties we use \verb+RibbonDB+ dataset\footnote{\url{http://solarmuri.ssl.berkeley.edu/~kazachenko/RibbonDB/}} \citep{Kazachenko2017}. \verb+RibbonDB+ database contains 3137 solar flare ribbon events corresponding to every flare of GOES class C1.0 and greater within $45^\circ$ { of} the central meridian, from 2010 April until 2016 April, observed by the {Atmospheric Imaging Assembly (AIA, \citealt{Lemen2012}) on board the SDO} in 1600\AA{} (see Eq. 7-11 in \citealt{Kazachenko2017}). To remind the reader, we sum up flux { of positive and negative magnetic polarities within flare ribbons} to find the unsigned ribbon reconnection flux (or AR unsigned flux):
\begin{equation}
\Phi_\mathrm{ribbon}=\Phi^+_\mathrm{ribbon}+\Phi^-_\mathrm{ribbon}=\frac{\Phi^\mathrm{+(I_6)}+\Phi^\mathrm{+(I_{10})}}{2}+\frac{\Phi^\mathrm{-(I_6)}+\Phi^\mathrm{-(I_{10})}}{2}=\frac{\Phi^\mathrm{(I_6)}+\Phi^\mathrm{(I_{10})}} {2},
\label{rbneq}
\end{equation}
where  $|\Phi^+_\mathrm{ribbon}|$ and $|\Phi^-_\mathrm{ribbon}|$ are signed reconnection fluxes in each polarity. Above, we take an average between ribbon fluxes { in areas with} $c=6$ and $c=10$ times the median background intensity for the minimum ribbon brightness. We do { this} to account for the uncertainty in ribbon edge identification due to variable choice of background cutoff  
 (see Equations (5) and (6) in \citealt{Kazachenko2017}):
\begin{equation}
\Phi^\mathrm{\pm(I_c)}_\mathrm{ribbon}=\int_{I_c,(> or < 100G)} |B_n| dS.
\label{rbneqc}
\end{equation}
The error in the unsigned reconnection flux (or the ribbon area) is then
\begin{equation}
\Delta \Phi_\mathrm{ribbon}=\frac{\Phi^\mathrm{(I_6)}-\Phi^\mathrm{(I_{10})}} {2}.
\label{rbneq_err}
\end{equation}
Following \citet{Kazachenko2017}, we define percentages of the ribbon-to-AR magnetic fluxes and ribbon-to-AR areas, respectively, {as}
\begin{equation} \label{eqratios}
    R_{\Phi} = \frac{\Phi_{\mathrm{ribbon}}}{\Phi_{\mathrm{AR}}} \times100\%,
    \;\;\;\;\; R_{S} = \frac{S_{\mathrm{ribbon}}}{S_{\mathrm{AR}}}\times100\%.
\end{equation}
We also find mean magnetic flux density within AR and ribbon areas as $\bar{B}_\mathrm{AR}=\frac{\Phi_\mathrm{AR}}{S_\mathrm{AR}}$ and $\bar{B}_\mathrm{ribbon}=\frac{\Phi_\mathrm{ribbon}}{S_\mathrm{ribbon}}$, respectively.

\begin{table}
    \caption{List of variables {we analyzed for \nflares~$>$M1.0-class flares and \ncflares~$>$C5.0-class flares (see Appendix~\ref{app}) and reference to their data source: RibbonDB \citep{Kazachenko2017}, \citet{Plutino2023}, TEBBS  \citep{Sadykov2019} and \citet{Li2021}. The data is available \href{http://solarmuri.ssl.berkeley.edu/~kazachenko/SolarErupDB/}{online}.}}
    \begin{center}
    \begin{tabular}{ l  l r  r}
\toprule
 \bf{Variable}        & \bf{Description} & Units & Data Source Reference\\
 \midrule
$I_{\rm X,peak}$        &  Peak 1-8\AA{} X-ray flux, GOES & [W~m$^{-2}$] & \citet{Plutino2023} \\
$t_{\rm start}$        & Flare start time  &[UT] &  \citet{Plutino2023} \\
 $t_{\rm peak}$    & Flare peak time &[UT] &   \citet{Plutino2023}\\
 $t_{\rm end}$    & Flare end time  &[UT] &  \citet{Plutino2023}\\
 $\tau=t_{\rm end}-t_{\rm start}$    &  Flare duration & [min] &   \citet{Plutino2023}\\
 $\tau_\mathrm{rise}=t_{\rm peak}-t_{\rm start}$    & Flare duration (rise)  &[min] &   \citet{Plutino2023}\\
 $F_\mathrm{X}$    &  X-ray total flux (fluence) & [W~m$^{-2}s$]  &   \citet{Plutino2023}\\
$T_{\rm GOES}$        &  Flare peak temperature, GOES & [MK] & \verb+TEBBS+ \\
$EM_{\rm GOES}$        & Flare peak EM, GOES & [$10^{48}cm^{-3}$] &  \verb+TEBBS+\\
 $lon, lat$                & Flare longitude and latitude & [deg] &  \verb+RibbonDB+ \\
 $AR_{\rm number}$    & AR NOAA number &$-$&  \verb+RibbonDB+ \\
 $\Phi_{AR}$             &  Unsigned AR flux& [Mx] &  \verb+RibbonDB+ \\
 $\Phi_\mathrm{ribbon}$ & Unsigned reconnection flux & [Mx] &  \verb+RibbonDB+ \\
 $\Delta \Phi_{\rm ribbon}$ & Uncertainty in $\Phi_\mathrm{ribbon}$&
 [Mx] &  \verb+RibbonDB+ \\
 $\Phi_{\rm ribbon,imb}$ & Imbalance in $\Phi_\mathrm{ribbon}$& [Mx] &  \verb+RibbonDB+ \\
 $\dot \Phi_{\rm ribbon}$ & Peak reconnection flux rate & [Mx/s] &  \verb+RibbonDB+ \\
$\Delta \dot \Phi_{\rm ribbon}$ & Uncertainty in $\dot \Phi_{\rm ribbon,peak}$& [Mx/s] &  \verb+RibbonDB+ \\
$\dot \Phi_{\rm ribbon,imb}$ & Imbalance in $\dot \Phi_{\rm ribbon,peak}$  & [Mx/s] &  \verb+RibbonDB+ \\
$S_{\mathrm{AR}}$     & AR area  & [cm$^2$] &  \verb+RibbonDB+ \\
$S_{\mathrm{ribbon}}$ & Ribbon area & [cm$^2$] &  \verb+RibbonDB+ \\
 $\Delta S_{\rm ribbon}$    & Uncertainty in $S_{\mathrm{ribbon}}$ & [cm$^2$] &  \verb+RibbonDB+ \\
 $R_{\Phi}$             &  Reconnection flux fraction & [\%]&  \verb+RibbonDB+ \\
 $R_{S}$             & Ribbon area fraction & [\%] &  \verb+RibbonDB+ \\
$\epsilon$             &  Eruptivity: 1 - eruptive, 0 - confined & $-$ & \citet{Li2021} \\
\bottomrule
\end{tabular}
\label{vars_tab}
\end{center}
\footnotesize{$^a$ \url{http://solarmuri.ssl.berkeley.edu/~kazachenko/SolarErupDB/}}
\end{table}

We further expand the original \verb+RibbonDB+ dataset to include peak reconnection flux rate $\dot \Phi_{\rm ribbon,peak}$ and imbalances in reconnection flux and the peak reconnection flux rate, $\Phi_{\rm ribbon,imb}$ and $\dot \Phi_{\rm ribbon,imb}$. Peak reconnection flux rate here describes the global peak rate of magnetic reconnection flux over the whole active region.
The imbalance is the difference between positive and negative magnetic polarities $\Phi_\mathrm{imb}=\left|~|\Phi_\mathrm{+}|-|\Phi_\mathrm{+}|~\right|$ and should ideally be equal to zero for an isolated well-observed AR or a set of footpoints brightened by the flare ribbons. Therefore we use this quantity as an additional quality-control measure for our magnetic fluxes. Specifically, we define the imbalance in the reconnection flux as
\begin{equation}
\Phi_\mathrm{ribbon,imb}=\left|~|\Phi^+_\mathrm{ribbon}|-|\Phi^-_\mathrm{ribbon}|~\right|.
\label{rbn_imb}
\end{equation}
 
Similarly, we define imbalance in the peak reconnection flux rate {as}
\begin{equation}
\dot\Phi_\mathrm{ribbon,imb}=\left|~|\dot\Phi^+_\mathrm{ribbon}|-|\dot\Phi^-_\mathrm{ribbon}|~\right|,
\label{rbnrate_imb}
\end{equation}
where positive and negative values are signed peak reconnection rates in each polarity. {To minimize errors in the reconnection fluxes we restrict our analysis to events with reconnection flux imbalance between positive and negative polarities $\Phi_\mathrm{ribbon,imb}<20\%$. }

How suitable are ribbon reconnection flux and peak reconnection rate for description of release of magnetic energy in solar flares?   { As footpoints of reconnected fields, flare ribbons only identify magnetic fields participating in the flare reconnection. However, they do not directly capture dynamics of arcades adjacent or below reconnected fields that evolve as part of the overall relaxation of the coronal field \citep{Hudson2000,Wang2018,Yadav2023}. What is the energetic significance of changes in fields that do not directly participate in the reconnection process? From comparison of magnetic energy released in the reconnection process and CMEs' energies (summed kinetic and gravitational potential energies), \citet{Zhu2020} found that for large M- and X-class flares reconnection dominates CME acceleration in fast CMEs, in agreement with earlier works by \citet{Longcope2007,Kazachenko2012}. However, for smaller B- and C-class flares, work done by the reconnection electric field in the current sheet itself might not be enough to fuel the eruption. Furthermore, there is a strong correlation between the reconnection flux and the GOES peak X-ray flux ($r_s=[0.6-0.9]$, \citealt{Sindhuja2020,
Toriumi2017, Kazachenko2017,Chernitz2018, Kazachenko2022r}). Based on these arguments, we assume that it is reasonable to expect that the total amount of energy
released in an event {\em scales with} magnetic reconnection flux and
reconnection rate, recognizing that these observables capture the key
aspects 
of the relaxation process. }

{To find the flare GOES peak X-ray flux and fluence, and also flare start, peak and end times, $I_\mathrm{X,peak}, F_\mathrm{X}, t_\mathrm{start}, t_\mathrm{peak}, t_\mathrm{end}$, we use a database by \citet{Plutino2023}\footnote{https://github.com/nplutino/FlareList}. In contrast to Heliophysics Event Catalog (HEC) that we used in  \verb+RibbonDB+ catalog to identify flare peak X-ray flux, start, peak and end times, \citet{Plutino2023} catalog uses a method by \citet{Aschwanden2012} to identify flares in the GOES SXR light curve. As a result, { the catalog of} \citet{Plutino2023} has slightly different values for peak X-ray fluxes and also flare times, compared to \verb+RibbonDB+ (see Figure~7 in \citealt{Plutino2023}). 
For flare total and rise time durations we use
\begin{equation}
\tau=t_\mathrm{end}-t_\mathrm{start} 
    \;\;\;\;\; and     \;\;\;\;\;  \tau_\mathrm{rise}=t_\mathrm{peak}-t_\mathrm{start},
\end{equation}
respectively.
In our analysis, we only used \verb+RibbonDB+ flare peak X-ray fluxes for initial selection of events.}

To find the GOES peak temperature $T_\mathrm{GOES}$ and emission measure $EM_\mathrm{GOES}$,
we use Temperature and EM-Based Background Subtraction algorithm (\verb+TEBBS+) dataset\footnote{https://github.com/vsadykov/TEBBS} from \citet{Sadykov2019}, deduced from GOES SXR observations. In { the} {\verb+TEBBS+ dataset \citet{Sadykov2019} apply \verb+TEBBS+ algorithm  \citep{Bornmann1990,Ryan2012}} to estimate  $T_\mathrm{GOES}$ and $EM_\mathrm{GOES}$ for all solar flares from 2002 January until 2017 December. $T_\mathrm{GOES}$ describes the maximum temperature of the dense chromospheric plasma after the flare starts.  $EM_\mathrm{GOES}$ describes the maximum emission measure that the chromospheric plasma reaches as it first evaporates and then condenses back into the chromosphere. Note that in this approach, as the GOES data includes only two SXR energy channels, both the flare temperature and EM are calculated in a single-temperature approximation, {which is a simplification given} evidence for multi-temperature structure \citep{Warmuth2016}.


In addition, {we use } GOES peak temperature  $T_\mathrm{GOES}$ and emission measure  $EM_\mathrm{GOES}$ {to} calculate flare thermal energy at the time of peak temperature, $E_\mathrm{th,GOES}=3 n k_B T_\mathrm{GOES} V$.  From the emission measure $EM_\mathrm{GOES} \approx n^2 V$, we find density $n=\sqrt{\frac{EM_\mathrm{GOES}}{V}}$, where for the flare {loop} volume $V$ we use $V=(\frac{S_\mathrm{ribbon}}{2})^{3/2}$. Then the expression for thermal energy becomes \citep{Reep2019}
\begin{equation}
E_\mathrm{th,GOES}=\frac{3}{8^{1/4}}k_B (EM_\mathrm{GOES})^{1/2} S^{3/4}_\mathrm{ribbon}T_\mathrm{GOES}.
\label{eth}
\end{equation}

Finally, to define whether each flare is confined or {eruptive,} we use { the database of} \citet{Li2021}\footnote{https://nadc.china-vo.org/res/r101068/}. This database includes 719 flares ($251$ eruptive and $468$ confined) of GOES class C5.0 and greater that occurred within $45^\circ$ { of} the central meridian, from June 2010 until June 2019. To define flare eruptivity, this dataset uses the CDAW CME catalog\footnote{https://cdaw.gsfc.nasa.gov/CME\_list/} \citep{Yashiro2004} based on the Solar and Heliospheric Observatory/Large Angle and Spectrometric Coronagraph (SOHO/LASCO, \citealt{Brueckner1995}) observations, {data from the twin Solar Terrestrial Relations Observatory (STEREO) and EUV wave detections from the AIA.} 

\begin{figure*}[tb!]
    \begin{center}
     \centerline{
    \includegraphics[width=1.0\textwidth]{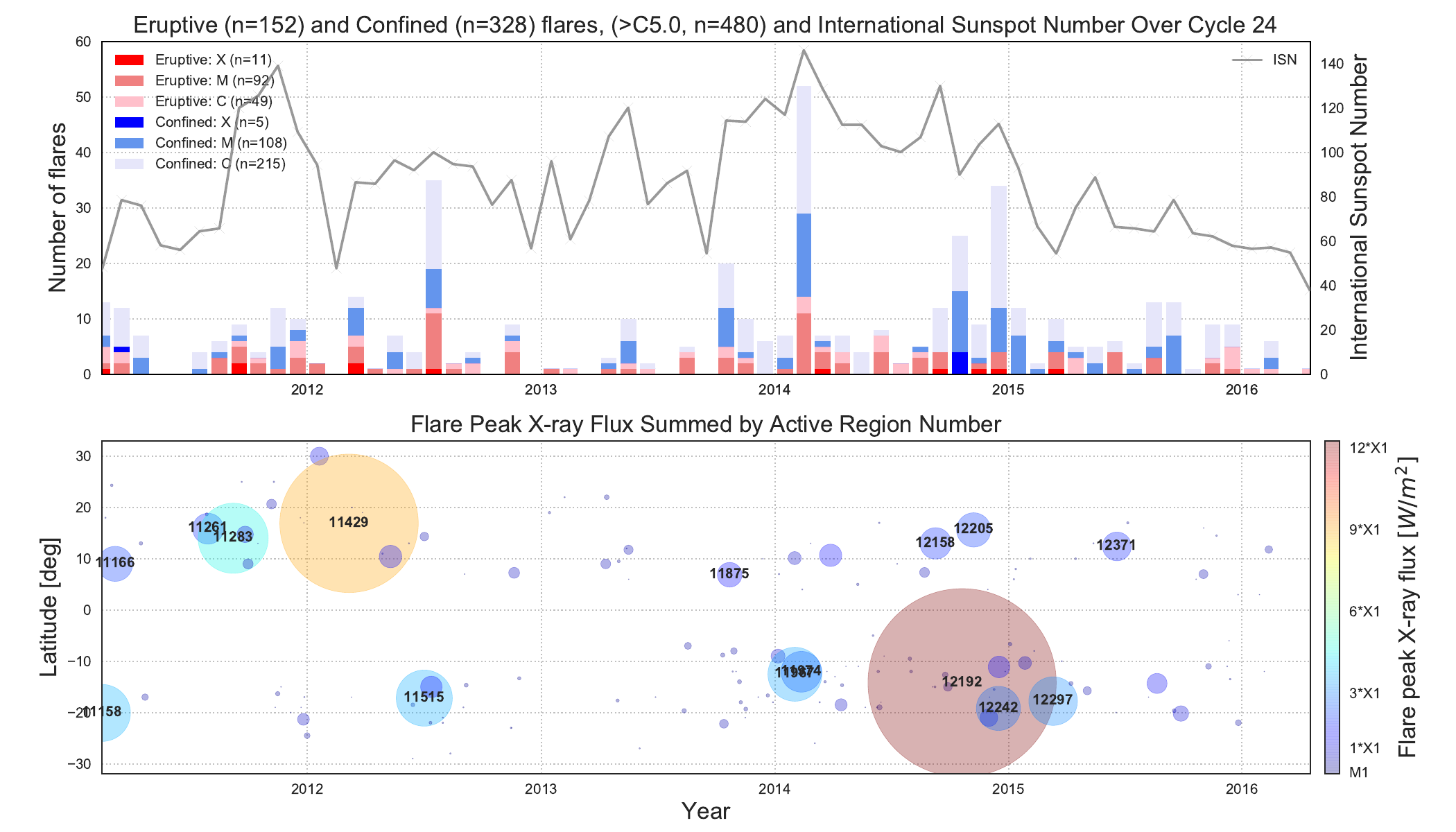} 
    }
     \caption{Our dataset of {$480$ flares ($152$ eruptive and $328$ confined flares), including {$103$} eruptive and {$113$} confined M- and X-class flares (see \S\ref{results}) and $49$ eruptive and $215$ confined C-class flares (see Appendix~\ref{app}). Top panel: number of confined and eruptive C-, M- and X-class flares each month analyzed in this paper (left axis) and sunspot number from 2010 April until 2016 April (right axis). These flares correspond to all events above C5.0 from the flare ribbon database RibbondDB.}  Bottom panel: flare peak X-ray flux and { latitude} grouped by AR vs. time; circle size and color correspond to the peak X-ray flux summed over each AR number. See \S\ref{methodology} for details.}
 \label{fig01} 
\end{center}
\end{figure*}

Figure~\ref{fig01} shows our final dataset {of {$480$} flares, $152$ eruptive and $328$ confined, with GOES class C5.0 and above, that we use for analysis of large ($>M1.0$) and all ($>C5.0$) flares. The top panel shows the total number of  confined and eruptive flares as a function of time; the bottom panel shows their location on the solar disk grouped by ARs and colored by the cumulative GOES peak X-ray class.}  
As the fraction of confined flares increases {for} smaller GOES peak X-ray class, we intentionally limit our first study to flares above { M1.0, which} {includes a roughly} equal number of confined and eruptive flares:
 {$103$} eruptive and {$113$} confined flares. This list corresponds to {\it all flares of GOES class M1.0 and above} observed by SDO from $2010$ to $2016$ April within $45^\circ$ { of} the central meridian, spanning the majority of solar cycle $24$. In this final dataset we determine {flare GOES X-ray properties { using the dataset of} \citet{Plutino2023} dataset,} flare ribbon properties from the \verb+RibbonDB+ dataset \citep{Kazachenko2017}, flare thermodynamic properties using { the} \verb+TEBBS+ dataset \citep{Sadykov2019} and flare eruptivity using \citet{Li2021} list. 

{ We describe the quantitative relationship between flare and AR properties, e.g., $\mathbb{X}$ and $\mathbb{Y}$, using the Spearman ranking correlation coefficient, $r_s(\mathbb{X},\mathbb{Y})$.  Specifically for qualitative strength of the correlation $r_s$ we use}:
$r_s \in [0.2,0.39]$ -- weak,
$r_s \in [0.4,0.59]$ -- moderate,
$r_s \in [0.6,0.79]$ -- strong, and
$r_s \in [0.8,1.0]$ -- very strong.

\subsection{Uncertainties}\label{unc}
{We recognize that the variables we analyze have} uncertainties. For example, identification of eruptive/confined flares might be ambiguous due to observational constraints (see e.g., an $M4.0$ flare on 2011 September 26 in AR 11302 that was identified as eruptive { in the catalog of} \citet{Li2021}, but confined by \citealt{Gupta2021}). {As} another example, reconnection fluxes in the \verb+RibbonDB+ using 1600 \AA{} SDO images are slightly different from reconnection fluxes evaluated using Kanzelhohe Solar Observatory (KSO) ribbon images in  $H_\alpha$ and $CaK$ bands \citep{Sindhuja2019}. While we could not re-examine all events, where possible, we evaluated uncertainties without modifying the main quantities. For unsigned reconnection fluxes $\Phi_\mathrm{ribbon}$ (Eq.~\ref{rbneq}), we evaluated errors based on uncertainty in the median background intensity $\Delta \Phi_\mathrm{ribbon}$ (Eq.~\ref{rbneq_err}). For unsigned peak reconnection fluxes and their rates, we evaluated error proxies based on the imbalance over positive and negative polarities (Eqs.~\ref{rbn_imb} and \ref{rbnrate_imb}). {Furthermore we restrict our analysis to events with reconnection flux imbalance of less than $20\%$. Note that we have not assessed the accuracy of \verb+TEBBS+ estimates for peak temperature and emission measure and confinement/eruptivity detections { of \citet{Li2021}}.}


\section{Results: Flares of GOES Class M1.0 and Larger}\label{results}
%
 \begin{figure*}[!tb]
    \includegraphics[width=1.0\textwidth]{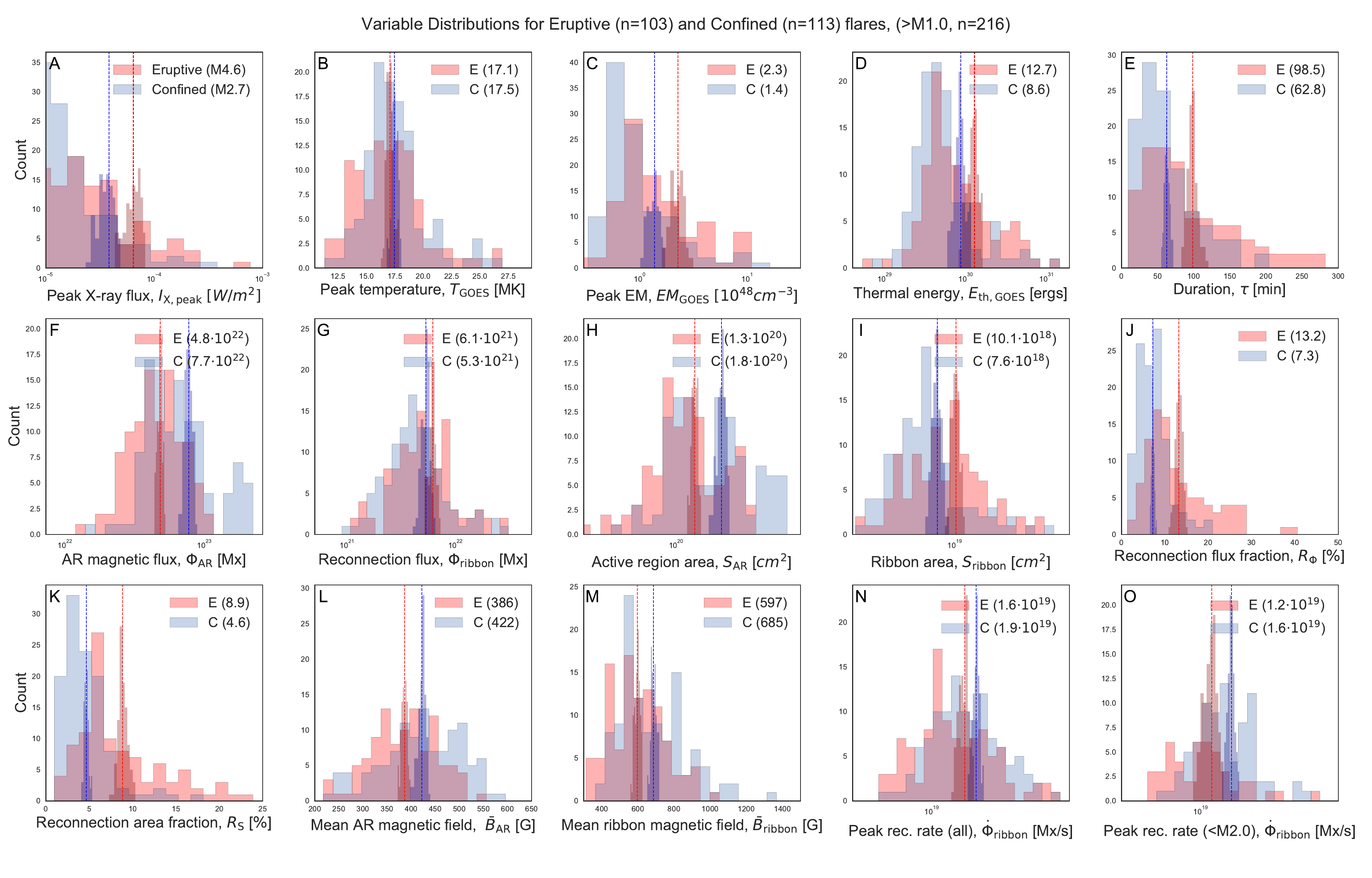}  
    \caption{{Results: Distributions of various physical variables for eruptive (red) and confined (blue) events ($>M1.0$ GOES flare class). Vertical dotted lines and numbers in the top right corner indicate the mean {value for} each variable within eruptive and confined flare groups. Darker-color histograms near dotted lines show distributions of the mean, if we select $100$ random subsamples of the half of the events in each sample, and reflect variability of the mean due to {event selection. See \S\ref{results_dist} for details.} }}
 \label{fig:dist}
 \end{figure*}
 
\begin{table*}[tbh!]
\caption{Confined vs. Eruptive flares: comparison of typical ranges and mean values of their magnetic and thermodynamic properties {for \nflares~$>$M1.0-class flares}. The typical range of each quantity is described as the $20^{th}$ to $80^{th}$ percentile $\mathbb{X}[P_{20},P_{80}]$.  {See \S\ref{results_dist} for details and Fig.~\ref{fig:dist} for  distribution plots of all variables}.} 
\begin{center}
\begin{tabular}{l l l l l l l}
\toprule
 & & \multicolumn{2}{c}{\textbf{CONFINED~FLARES}} & \multicolumn{2}{c}{\textbf{ERUPTIVE~FLARES}} \\ 
  \cmidrule(r){3-4}    \cmidrule(r){5-6}
Quantity & Units & Typical range & Mean Value & Typical range & Mean Value & Figure \\
 $\mathbb{X}$ & & $\mathbb{X}_\mathrm{conf}[P_{20},P_{80}]$ & $\bar{\mathbb{X}}_\mathrm{conf}$ & 
 			      $\mathbb{X}_\mathrm{erup}[P_{20},P_{80}]$ & $\bar{\mathbb{X}}_\mathrm{erup}$ & \\
\midrule
Peak X-ray flux, $I_\mathrm{X,peak}$ & Flare class & [M1.4,M4.1] & $M2.7$ & $ [M1.8,M8.9]$ & $ M4.6$ & Fig.~\ref{fig:dist}(A) \\
Duration, ${\tau}$ &  $min$ &  $[31,91]$ & $63$ & $ [41,148]$ & $99$ & Fig.~\ref{fig:dist}(E) \\
Rise duration, ${\tau_\mathrm{rise}}$ &  $min$ &  $[8.2,21]$ & $16$ & $ [9.1,28.8]$ & $22$ & -- \\
Peak temperature, $T_\mathrm{GOES}$ & $MK$ & [15.6,18.6] & $17.5$ & $ [14.2,19.0]$ & $ 17.1$ & Fig.~\ref{fig:dist}(B) \\
Peak EM, $EM_\mathrm{GOES}$ & $10^{48}~cm^{-3}$ & [0.5,1.5] & $1.4$ & $ [0.7,3.2]$ & $ 2.3$ & Fig.~\ref{fig:dist}(C) \\
Peak thermal energy, $E_\mathrm{th,GOES}$ & $10^{29}~ergs$ & [2.8,8.4] & $8.6$ & $ [3.8,15.3]$ & $ 12.7$ & Fig.~\ref{fig:dist}(D) \\
AR mag. flux, $\Phi_\mathrm{AR}$ &  $10^{21}~Mx$ &  $[40,95]$ & $77$ & $ [30,63]$ & $48$ & Fig.~\ref{fig:dist}(F) \\
Reconnection mag. flux, $\Phi_\mathrm{ribbon}$ &  $10^{21}~Mx$ &  $[2.6,6.4]$ & $5.3$ & $ [2.8,8.1]$ & $6.1$ & Fig.~\ref{fig:dist}(G) \\
AR area, $S_\mathrm{AR}$ &  $10^{18}~cm^2$ &  $[98,243]$ & $180$ & $ [83,177]$ & $131$ & Fig.~\ref{fig:dist}(H) \\
Ribbon area, $S_\mathrm{ribbon}$ &  $10^{18}~cm^2$ &  $[4.2,8.7]$ & $7.6$ & $ [5.1,13.5]$ & $10.1$ & Fig.~\ref{fig:dist}(I) \\
Rec. flux fraction, $R_\Phi$ &  $10^{22}~Mx$ &  $[4.4,9.2]$ & $7.3$ & $ [7.2,19.2]$ & $13.2$ & Fig.~\ref{fig:dist}(J) \\
Rec. area fraction, $R_S$ &  $10^{22}~Mx$ &  $[2.6,6.1]$ & $4.6$ & $ [4.7,12.7]$ & $8.9$ & Fig.~\ref{fig:dist}(K) \\
Mean AR mag. field, $\bar{B}_\mathrm{AR}$ &  $G$ &  $[348,539]$ & $422$ & $ [329,448]$ & $386$ & Fig.~\ref{fig:dist}(L) \\
Mean ribbon mag. field, $\bar{B}_\mathrm{ribbon}$ &  $G$ &  $[539,821]$ & $685$ & $ [454,707]$ & $597$ & Fig.~\ref{fig:dist}(M) \\
Peak reconnection rate, $\dot\Phi_\mathrm{ribbon}$ &  $[10^{19}~Mx/s]$ &  $[1.1,2.4]$ & $1.9$ & $ [0.8,2.2]$ & $1.6$ & Fig.~\ref{fig:dist}(N) \\
\bottomrule
%
\end{tabular}
\label{meantab} 
\end{center} 
\end{table*}

In Figures~\ref{fig:dist}-\ref{fig:scat} we show the main results of this paper: distributions, {Spearman correlation matrix} and scatter plots of variables listed in Table~\ref{vars_tab} for flares of GOES class M1.0 and larger. In Table~\ref{meantab} we show typical ranges and mean values for all variables. 

\subsection{Results: distributions of thermodynamic and magnetic properties in confined and eruptive flares}\label{results_dist}
In Figure~\ref{fig:dist}(A-E) we first analyze the {\it distributions of thermodynamic properties}: GOES peak X-ray flux, peak temperature, emission measure, thermal energy and flare duration. In Table~\ref{meantab} we compare {variables'} typical range and mean values {within confined and eruptive samples}.

{From Figure~\ref{fig:dist}A we} see that while the total number of confined and eruptive flares above M1.0 is similar {($n_\mathrm{erup}=103$ vs. $n_\mathrm{conf}=113$)}, with increasing peak X-ray flux the number of confined flares vs. eruptive flares gradually decreases. 
{As a result, in our sample eruptive flares have higher mean peak X-ray flux than eruptive flares: M4.6 vs. M2.7, respectively.} 
Similarly, {the typical ranges of both peak emission measure and thermal energy are} higher in eruptive flares (Figures~\ref{fig:dist}C and~\ref{fig:dist}D):
 { 
\begin{equation}
EM_\mathrm{GOES,erup}[P_{20},P_{80}]= [0.7,3.2]\times10^{48}cm^{-3}
    \;\;\;\;\; vs.     \;\;\;\;\;  EM_\mathrm{GOES,conf}[P_{20},P_{80}]= [0.5,1.5]\times10^{48}cm^{-3},
\end{equation}
and
\begin{equation}
E_\mathrm{th,GOES,erup}[P_{20},P_{80}]= [3.8,15.3]\times10^{29}ergs
    \;\;\;\;\; vs.     \;\;\;\;\;  E_\mathrm{th,GOES,conf}[P_{20},P_{80}]= [2.8,8.4]\times10^{29}ergs.
\end{equation}
}
In contrast, the mean SXR temperature is slightly lower in eruptive than confined flares (see Figure~\ref{fig:dist}B). Lower peak temperatures and higher emission measures of eruptive flares {are} consistent { with case studies, by \citet{Qiu2022},} of $3$ eruptive and $3$ confined flares, where they found the high-temperature component at the flare onset in confined events in spite of higher {flare classes of eruptive events in their sample. Similarly, \citet{Kahler2022} analyzed hundreds of flares, finding that for a given X-ray peak flux, confined flares have higher temperatures than eruptive flares confirming the results of \citet{Kay2003}.} 

We find that confined flares {have shorter durations} than eruptive events, with the mean of {$\tau_\mathrm{conf}=63\pm4$ minutes and  {$\tau_\mathrm{erup}=99\pm7$} minutes, and the typical range, as the $20^\mathrm{th}$ to $80^\mathrm{th}$ percentile, of $[31,91]$ minutes and $[41,148]$ minutes, respectively} (see Table~\ref{meantab}). This finding is consistent with {e.g.} \citet{Webb1987,Toriumi2017,Hinterreiter2018}, who reported that eruptive flares {last} longer than confined ones.

In Figure~\ref{fig:dist}(F-O) we analyze the {\it distributions of magnetic properties}. 
We find that confined flares have larger AR magnetic fluxes and areas (Figure~\ref{fig:dist}F and Figure~\ref{fig:dist}H). {Confined flares} have also larger mean magnetic field strengths within ribbon and AR areas  (Figure~\ref{fig:dist}L and Figure~\ref{fig:dist}M).
{The above differences are} consistent with magnetic cage interpretation, {where ARs with large magnetic flux have confined eruptions.}
 Practically, this means that for {\it large flares of GOES class M1.0 and above:} larger ARs, especially those with unsigned magnetic flux of $10^{23}$ Mx and {above,} tend to host confined flares; medium ARs with $3\times10^{22} Mx<\Phi_\mathrm{AR}<10^{23}$ Mx host both confined and eruptive flares; smaller ARs with $\Phi_\mathrm{AR}<3\times10^{22}$ Mx tend to host eruptive flares. {Note that if we include smaller C-class flares, then we find that smaller active regions host a similar number of confined and eruptive flares (see Figure~\ref{fig:dist_small}F in Appendix~\ref{app}).}
 
{From ribbon analysis,} we find that the {magnetic} reconnection fluxes have similar distributions in both samples (Figure~\ref{fig:dist}G), while the ribbon areas are larger for eruptive flares (Figure~\ref{fig:dist}I). As a result, both reconnection flux and area fractions are larger for eruptive flares. { These results are consistent with a scenario { in which} weaker strapping field makes flares more eruptive  (Figure~\ref{fig:dist}J and Figure~\ref{fig:dist}K and also earlier works by e.g., \citealt{Toriumi2017})}. {Another factor that could explain why larger ARs host more confined flares is the relationship between AR size and access to periphery. Smaller regions have larger AR area fractions near their peripheries, with potentially better access to global field. In contrast, larger ARs have smaller AR area fractions near peripheries and hence tend to be more confined. Furthermore, the concept of closeness to periphery could explain why eruptive flares tend to have weaker magnetic fields, since the magnetic fields on { AR peripheries are generally} weaker. In contrast, confined flares are further from the periphery, where magnetic field is stronger, therefore mean magnetic field within ribbons for confined flares is stronger.}
{The above results of larger} magnetic field and {smaller} ribbon area within confined flares imply that in confined flares reconnection  proceeds in more compact current sheets at lower coronal heights with larger field strengths. On the other hand, decreased magnetic field and increased ribbon area within eruptive flares implies that in eruptive flares reconnection  proceeds in {more extended} current sheets at larger coronal heights with smaller magnetic field strengths. 

Finally, we find that  peak reconnection rates are {slightly} higher in confined vs. eruptive flares (Figures~\ref{fig:dist}N and ~\ref{fig:dist}O). In other words, the same amount of reconnection flux, which we find to be very similar for eruptive and confined events in Figure~\ref{fig:dist}G, gets reconnected at higher rates 
in confined events compared to eruptive flares
(see Figure~\ref{fig:dist}E). We hypothesize that the fact that confined flares are more compact and as a result occur lower in the corona involving stronger magnetic fields is the main factor that leads to increased peak reconnection rates in confined flares. {At first,} these results for the reconnection rates {might} seem different from \citet{Hinterreiter2018} {and }\citet{Chernitz2018}, who analyzed local peak reconnection flux rates (i.e. electric fields $E=v_\mathrm{rbn}B_\mathrm{n}$) and {global} peak reconnection flux rates in $50$ flares, finding that both are similar or higher in the eruptive flares compared to confined flares. We {note, however}, that our results here are for large flares above M1.0 with similar distributions of flare classes within confined and eruptive samples. In contrast, \citet{Hinterreiter2018} {and} \citet{Chernitz2018} samples are very different, {with flare class ranging from B to X-class flares and} confined sample being much  {weaker than the eruptive} one. 
Accounting for sample differences, we suggest that the two results would be consistent, i.e. a confined flare with similar reconnection and peak X-ray fluxes as an eruptive flare, would tend to have a higher peak reconnection rate {than} the eruptive flare. {Figure~\ref{fig:dist}(C) and Figure 7 in \citet{Chernitz2018} confirm our guess, showing that for the same peak X-ray flux confined flares have higher peak reconnection rate.}
 
\subsection{Results: relationship between thermodynamic and magnetic properties in confined and eruptive flares}\label{results_scatter}
%
\begin{figure*}[!tb]
    \includegraphics[width=0.6\textwidth]{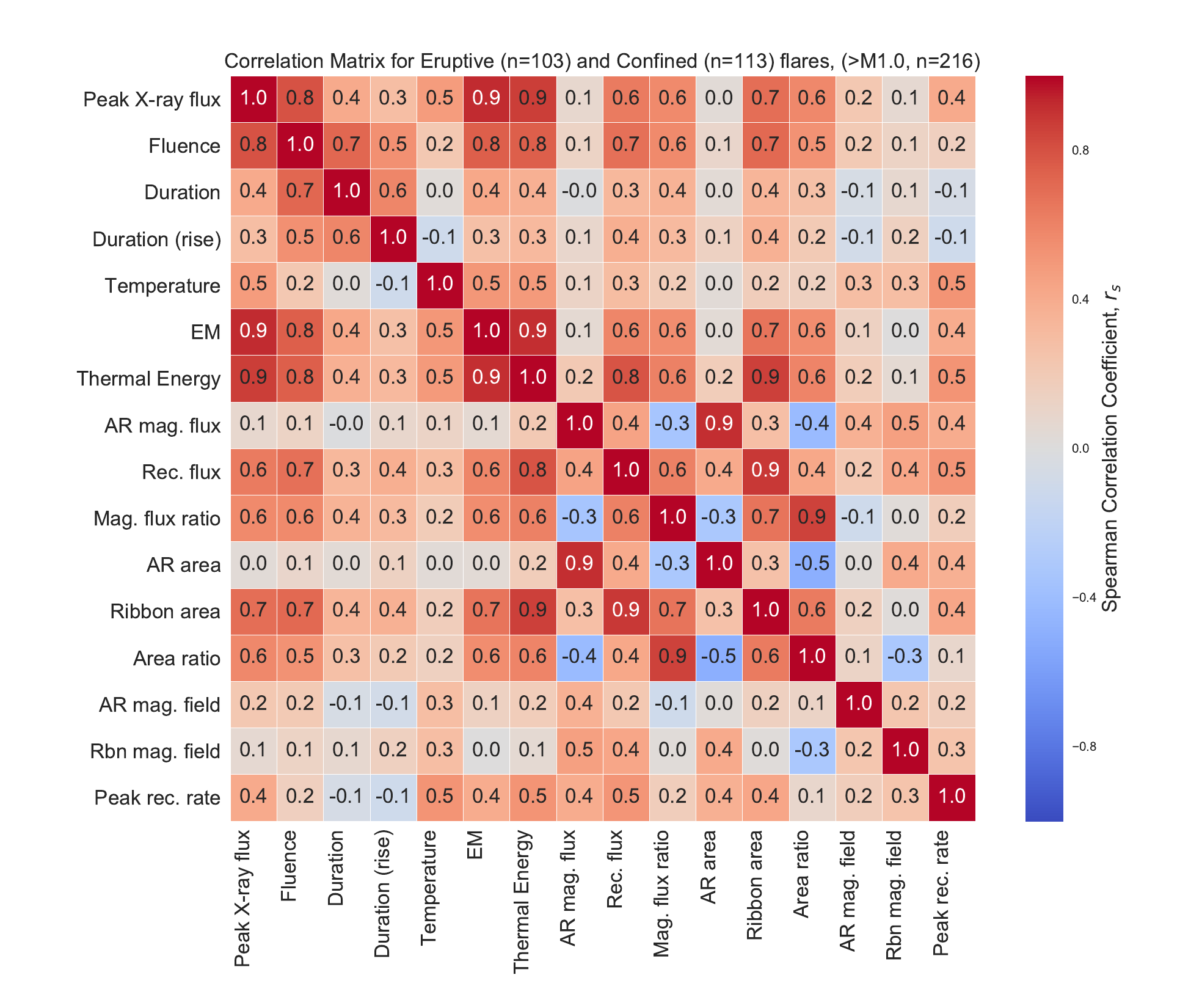}
    \caption{{Results: Correlation matrix showing Spearman correlation coefficient, $r_s$, between different physical variables for all flares with GOES flare class $>M1.0$. We describe the qualitative strength of the correlation using the following guide:  
 $r_s \in [0.2,0.39]$ -- weak,
$r_s \in [0.4,0.59]$ -- moderate,
$r_s \in [0.6,0.79]$ -- strong, and
$r_s \in [0.8,1.0]$ -- very strong.
}{See \S\ref{results_scatter} for details.} }
 \label{fig:corrmat}
 \end{figure*}
\begin{figure*}[!tb]
    \includegraphics[width=1.0\textwidth]{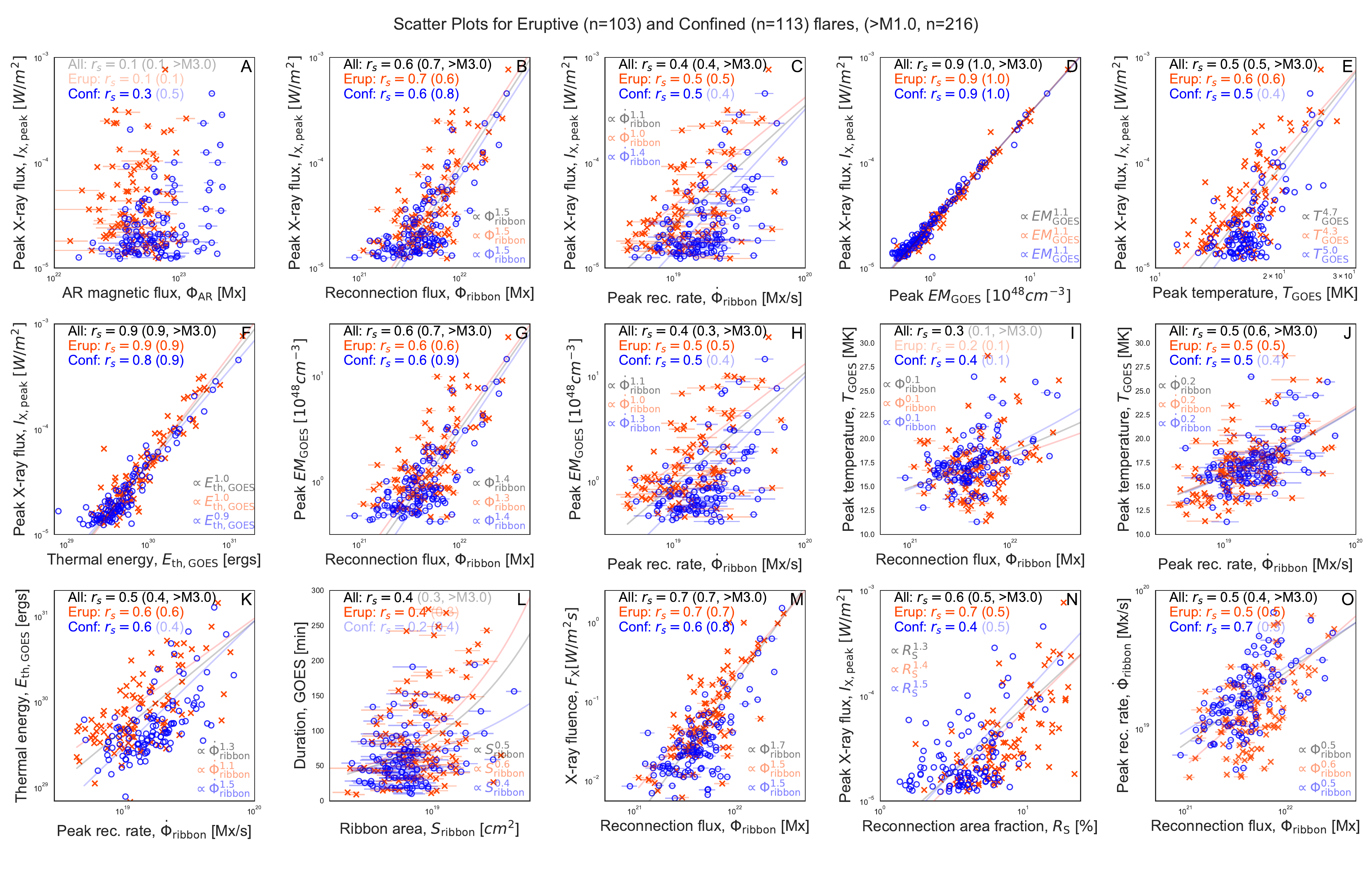}
    \caption{{Results: Scatter plots between different physical variables for confined (blue) and eruptive (red) events} ($>M1.0$ GOES flare class). In the corner we show Spearman correlation coefficient $r_s$ between each pair of plotted variables for all flares (black) and eruptive (red) and confined (blue) flares separately, {where bold and dim fonts indicate statistical significance of the correlation with p-values below and above $p=0.01$, respectively.} The values in the { parentheses} show values for flares above M3.0. {See \S\ref{results_scatter} for details.}}
 \label{fig:scat}
 \end{figure*}
{In Figure~\ref{fig:corrmat} we show correlation matrix with Spearman correlation coefficients between all physical variables that we analyze in this paper. We show this matrix for illustrative purposes, to describe the relationship between all variables, especially those that we do not show in the scatter plots  in Figure~\ref{fig:scat}.}
 
In Figure~\ref{fig:scat}(A-F) we first analyze how peak X-ray flux scales with other variables {and compare how confined and eruptive flare properties differ for a fixed peak X-ray flux}. 

The GOES peak X-ray flux has {a moderate to} high correlation with several thermodynamic properties, including peak temperature ({$r_s=0.5$},  Figure~\ref{fig:scat}E), emission measure ($r_s=0.9$, Figure~\ref{fig:scat}D) {and thermal energy ($r_s=0.9$, Figure~\ref{fig:scat}F)}. These correlations are in agreement with \citet{Feldman1996,Reep2019,Sadykov2019} and the standard chromospheric evaporation model \citep{Hirayama1974}, where GOES peak X-ray flux depends on the maximum EM.

The correlation between the AR magnetic flux and the peak X-ray flux is weak ($r_s=0.1$, Figure~\ref{fig:scat}A) -- larger active regions could host both large and small flares.
In contrast the correlation between the peak X-ray flux and reconnection flux is {strong} ($r_s=0.6$ to $0.7$ for flares above M3.0, Figure~\ref{fig:scat}B) in agreement with \citet{Kazachenko2017,Toriumi2017,Chernitz2018}, {meaning that} larger flares have larger amount of magnetic field participating in the reconnection. For flares above C1.0,  \citet{Kazachenko2017} {fit this relationship as $X_\mathrm{peak} \propto (\Phi_\mathrm{ribbon})^{1.5}$}, while \citet{Reep2019} find {similar} $X_\mathrm{peak} \propto (\Phi_\mathrm{ribbon]})^{1.6}$.  {Here we find $X_\mathrm{peak} \propto (\Phi_\mathrm{ribbon})^{1.5}$ for both confined and eruptive flares, consistent with \citealt{Kazachenko2017}.}

{Since the reconnection flux is calculated cumulatively by adding the newly reconnected flux, as flare progresses, it is physically more appropriate to compare reconnection flux with cumulative GOES SXR flux, like SXR fluence. In the past, SXR fluence has been widely used to compare flare and CME energetics (e.g., \citealt{Salas-Matamoros2015}). From Figure~\ref{fig:scat}M we find that the correlation between the SXR fluence and reconnection flux is strong, $r_s=0.7$, only slightly above the correlation coefficient between GOES peak X-ray flux and reconnection flux ($r_s=0.6$). This result agrees with the previous study of \citet{Sindhuja2019} who found that reconnection flux has stronger correlations with fluence than with the peak X-ray flux: $r_s=0.8$ vs $r_s=0.6$ in their sample}.
In addition, {we find that} the flare peak X-ray flux has moderate correlation with flare area ratio and flux ratio ({$r_s=0.6$}, Figure~\ref{fig:scat}N), in agreement with \citealt{Kazachenko2017}.

{We find that the} correlation between the peak X-ray flux and the peak reconnection flux rate is moderate, $r_s=0.4$ {for both smaller and larger flares above M3.0 (Figure~\ref{fig:scat}C), \bf which is significantly weaker than correlation between peak X-ray flux and reconnection flux ($r_s=0.6$)}. 
 
In Figures~\ref{fig:scat}(G-H) we quantify how emission measure is related to other reconnection flux and its peak rate. 
{As emission measure is very strongly correlated with the GOES peak X-ray flux ($r=0.9$), it has the same correlation coefficients with other variables as the GOES peak X-ray flux: strong correlations with reconnection flux ($r_s=0.6$, Figure~\ref{fig:scat}G), ribbon area ($r_s=0.7$), flux ratio and area ratio ($r_s=0.6$) and moderate correlations with GOES temperature ($r_s=0.5$)
and peak reconnection rate ($r_s=0.4$, Figure~\ref{fig:scat}H).}
%

{For the first time, in Figures~\ref{fig:scat}(I-J) we compare how GOES peak temperature is related to other reconnection properties.} We find that the peak temperature has very weak correlation with the reconnection flux ($r_s=0.3$), which disappears for stronger flares ($r_s=0.1$ for flares above M3.0, Figure~\ref{fig:scat}I). {In contrast to reconnection fluxes}, the peak temperature is {sensitive} to the peak reconnection rate with moderate {to strong correlation coefficients, $r_s=0.5$ and $r_s=0.6$, for flares above $M1.0$ and $M3.0$, respectively (Figure~\ref{fig:scat}J). }

How does thermal energy scale with the reconnection flux?  
We find a strong to very strong correlation between the two ($r_s=0.8$), indicating that reconnection flux defines the thermal energy output of the flare.  We find that the correlation between the thermal energy and the peak reconnection flux rate is weaker ($r_s=0.5$, Figure~\ref{fig:scat}K), implying that the total reconnection flux has a stronger effect on the thermal energy than the peak reconnection rate, similar to the effect on the emission measure.

We find that flare duration has a moderate correlation with the flare ribbon {area} ($r_s=0.4$, Figure~\ref{fig:scat}L). These results are consistent with statistical results by \citet{Toriumi2017,Reep2019} who found that in large flares above M5.0 flare durations scale with ribbon separation distances{, while in weaker flares the correlation is weak \citep{Reep2019}}. One possible interpretation is that flare duration could be defined by the Alfv\'{e}n transit time {across the flaring area, which is related to ribbons' size. In addition, {the correlation between} flare duration with flare ribbons size could be explained with the avalanche flare model (e.g., \citealt{Lu1991,Lu1995b}): if a flare proceeds as a chain reaction of one reconnection event triggering others, then reconnecting larger flux would take more time.} 


{ Finally, in Figure~\ref{fig:scat}O we look at the statistical relationship between the reconnection fluxes and peak reconnection flux rates. From this plot  we find  that the two variables are moderately correlated ($r_s=0.5$). We also find that for a fixed reconnection flux, confined flares  have higher peak reconnection rates.} 

{We then use scatter plots in Figure~\ref{fig:scat} to describe how variables within confined and eruptive samples differ for fixed flare class or emission measure.  
We find that for a fixed peak X-ray flux, confined flares have higher { reconnection rates and peak temperatures  (see Figures~\ref{fig:scat}C and E). For thermal energies, only large confined flares, of flare class X1.0 and above, have higher thermal energies than eruptive flares (see Figure~\ref{fig:scat}F).} 

 {These results are consistent with a qualitative scenario of energy partition between CME kinetic energy and flare thermal energy, where CME removes energy from the source region, leading to less thermal energy available for flares and lower peak flare temperatures in eruptive flares.}

{To improve our understanding of the underlying physical mechanisms driving the above correlations, we use a framework of {\it extensive and intensive flare properties} { \citep{Tan2007,Welsch2009,Bobra2016,Kazachenko2022}}. 

Extensive properties are those that correlate or scale with flare size, described by peak X-ray flux or X-ray fluence, while intensive properties are those that do not scale with flare size. 
Above, we find the following extensive flare properties with a strong correlation with flare fluence, $F_\mathrm{X}$: peak X-ray flux, emission measure, thermal energy, reconnection flux and ribbon area: $r_s(F_\mathrm{X}; [I_\mathrm{peak},EM_\mathrm{GOES},E_\mathrm{th,GOES},\Phi_\mathrm{ribbon},S_\mathrm{ribbon}])=[0.8,0.8,0.8,0.7,0.7]$. 
We also find that many extensive properties have a strong correlation with each other: e.g., $r_s(S_\mathrm{ribbon},[EM_\mathrm{GOES},E_\mathrm{th,GOES}])=[0.7,0.9]$. 

{Above,} we also find some variables that behave more like intensive properties, such as temperature and reconnection flux rate. These properties do not exhibit a correlation with X-ray fluence,  $r_s(F_\mathrm{X}; [T_\mathrm{GOES},\dot\Phi_\mathrm{ribbon},])=[0.2,0.2]$, and only moderate correlations with peak X-ray flux:  $r_s(I_\mathrm{X,peak}; [T_\mathrm{GOES},\dot\Phi_\mathrm{ribbon},])=[0.5,0.4]$. 
However, both temperature and reconnection rate exhibit moderate correlation with each other:  $r_s(T_\mathrm{GOES},\dot\Phi_\mathrm{ribbon})=0.5$.} 

What controls the plasma temperature and emission measure in flares? In \citet{Longcope2016,Longcope2018}, it is the reconnection rate that {affects} the maximum values of plasma properties. {From} our analysis we find that the temperature is more controlled by peak reconnection rate ($r_s=0.5$), while emission measure and thermal energy are more controlled by the reconnection flux ($r_s=0.6$ and $r_s=0.8$). {For comparison the correlation between the reconnection flux and reconnection flux rate is only moderate ($r_s=0.5$, Figure~\ref{fig:scat}O)}. These observational  inferences are consistent with {the framework of correlation between intensive and extensive flare properties above} and case studies of temporally resolved profiles of reconnection rates, temperature and emissions (see e.g., \citealt{Qiu2010,Qiu2022}). {For example}, \citet{Qiu2022} analyzed $3$ eruptive and $3$ confined flares, finding that in all flares temperature rises and peaks early or co-temporal with the reconnection rate rise and peak. 



In this section we focused on larger flares, of class $M1.0$, and above. Note that while variable distributions we find within confined and eruptive samples are not drastically different, the differences in the mean values are statistically persistent for different sub-samples (see darker-tone shaded histograms centered at dotted mean values in Figure~\ref{fig:dist}).
If we include flares $C5.0$ and above, {as we do in Appendix~\ref{app}}, we find similar physical conclusions in terms of flare magnetic and thermodynamic properties, with all the {mean quantities being smaller due to smaller flare sizes}. The main difference for flares within  $C5.0-M1.0$ class vs. flares $M1.0$ and above, {which} we analyzed above, is that the fraction of confined flares drastically increases for flare class below $M1.0$, with fractions of confined flares of $80\%$ and higher as we go to lower flare class (see Appendix~\ref{app} for analysis including flares $>C5.0$). On the other hand, if we restrict our analysis to very large flares above $M5.0$, we find similar physical conclusions as for smaller flares; the only difference is in the duration distributions: confined flares have similar durations as eruptive flares for GOES classes $>M5.0$ (cf., for flares $>M1.0$ confined flares are slightly shorter). 

\section{Conclusions}\label{conc}

 We have compared magnetic and thermodynamics properties of 216 {confined and eruptive flares {\it of GOES class M1.0 and above} (see Figures~\ref{fig:dist}-\ref{fig:scat}). In Appendix~\ref{app} we have also included smaller flares of GOES class C5.0 and above and analyzed properties in 480 flares}. 
 
What are the differences in {average} {\it magnetic properties  between large confined and eruptive flares}?  We find that:
\begin{itemize}
    \item {Eruptive flares  have the same amount of reconnected flux as confined flares. However, unlike confined flares, they occur in smaller ARs with weaker magnetic field strengths. Their flare ribbons have weaker mean magnetic field strengths. As a result, they reconnect larger fractions of AR magnetic flux than confined flares.  
    \item Confined flares occur in larger ARs with stronger mean magnetic field strengths compared to eruptive flares. Their ribbons are more compact, have stronger magnetic field strengths and reconnect smaller fractions of AR magnetic flux. Both of these results are consistent with earlier studies. }
    \item {For the first time,} we find that {for the same peak X-ray flux confined flares have higher} mean peak reconnection rates than eruptive flares (see Figure~\ref{fig:scat}C). This together with the fact that HXR (or microwave) emissions are temporally correlated with the global (see \citealt{Qiu2022} and references therein) and local reconnection rates \citep{Temmer2007,Naus2021}, implies that large, eruptive flares are less efficient in particle acceleration than confined flares, in agreement with direct particle measurements (see e.g., \citealt{Thalmann2015,Cai2021}). 
\end{itemize}

The above results support previously described main factors that define whether a flare would be confined or eruptive: (1) the strength of the overlying field, given by the AR flux, 
{(2) the ratio between the reconnected magnetic flux and the active region flux}
and (3) the strength of the reconnected field (mean ribbon magnetic field). 
Here we have not investigated {either} ribbon field non-potentiality {or} its proximity to the AR edge, but previous works have shown that these differ in confined and eruptive flares (see summary in \S\ref{intro}).

For flares of M-class and above, what are the differences in {\it thermodynamic plasma properties} between confined and eruptive flares? Our conclusions are as follows. 
\begin{itemize}
    \item Large {\it eruptive} flares have larger mean peak X-ray fluxes, peak emission measures and thermal energies. They have longer durations, and slightly smaller peak temperatures.
    \item Large {\it confined} flares have smaller mean peak X-ray fluxes, peak emission measures and thermal energies. They have shorter durations, and slightly larger peak temperatures.
    \item Why, for similar reconnection fluxes, {do} confined flares have larger temperatures and peak reconnection flux rates? We explain it with lower coronal altitudes and stronger reconnection fields in confined flares (since confined flares are more compact).
    \item Why, for similar reconnection fluxes, {do} confined flares tend to have shorter durations than eruptive flares? We explain it with the smaller physical size of confined flares (ribbon areas) and as a consequence shorter Alfv\'{e}n transit time over the {flaring area}, than in eruptive flares  \citep{Reep2019}.    
    \item We find that peak X-ray flux and emission measure are {more strongly} controlled by the reconnection fluxes than peak reconnection rates. 
   On the other hand, peak temperature is {more strongly} controlled by the peak reconnection rate than the reconnection flux.  This is consistent with temporally-resolved flare case studies that show temperature peaking early on together with the reconnection rate peak, while emission measure slowly rises to its maximum values as the cumulative reconnection flux increases (see e.g., \citealt{Qiu2022}).  
\end{itemize}

To summarize, our findings indicate that, in general, in eruptive flares, reconnection proceeds { more slowly} in larger current sheets higher in the corona where the coronal magnetic field is weaker. In contrast, in confined flares, reconnection proceeds faster in more compact current sheets lower in the corona where the coronal magnetic field is stronger. 
{ As higher reconnection rates lead to more accelerated ions and electrons, large confined flares could be more efficient in producing ionizing electromagnetic radiation than eruptive flares.
 Given that flare high-energy radiation could affect exoplanet habitability conditions \citep{Yamashiki2019,Airapetian2020}, we speculate that confined flares could be important in shaping exoplanet conditions.} This, together with a lack of CME detections on solar-like stars, would imply potentially a larger role of confined flares in exoplanet habitability. 

In this {section} we focused on large flares of class $M1.0$ and above to have a similar number of eruptive and confined flares to analyze. Extending our analysis to flares of class $C5.0$ and above, we end up with a much larger fraction of confined flares, but find similar physical differences between magnetic/thermodynamic properties of confined/eruptive flares, with smaller mean values. For the results of our analysis for smaller flares, C5.0 and above, see Appendix~\ref{app} below. 

The catalog is available \href{http://solarmuri.ssl.berkeley.edu/~kazachenko/SolarErupDB/}{online}  in IDL SAV and CSV file formats, and can be used for a wide spectrum of quantitative studies in the future, including case studies of the outliers.
\\
\\
We thank the reviewer for providing comments that have improved the manuscript. We thank the HMI team for providing us with the vector magnetic field SDO/HMI data.  We thank the AIA team, in particular Marc DeRosa, for providing us with the UV SDO/AIA data. We thank the Croom Team for stimulating discussions.  We thank the US taxpayers for providing the funding that made this research possible. We acknowledge support from NASA LWS NNH17ZDA001N, 80NSSC19K0070, NASA ECIP NNH18ZDA001N and NSF CAREER SPVKK1RC2MZ3.

\appendix
\section{Expanding analysis from flares above M1.0 to flares above C5.0}\label{app}
 \begin{figure*}[!tb]
    \includegraphics[width=1.0\textwidth]{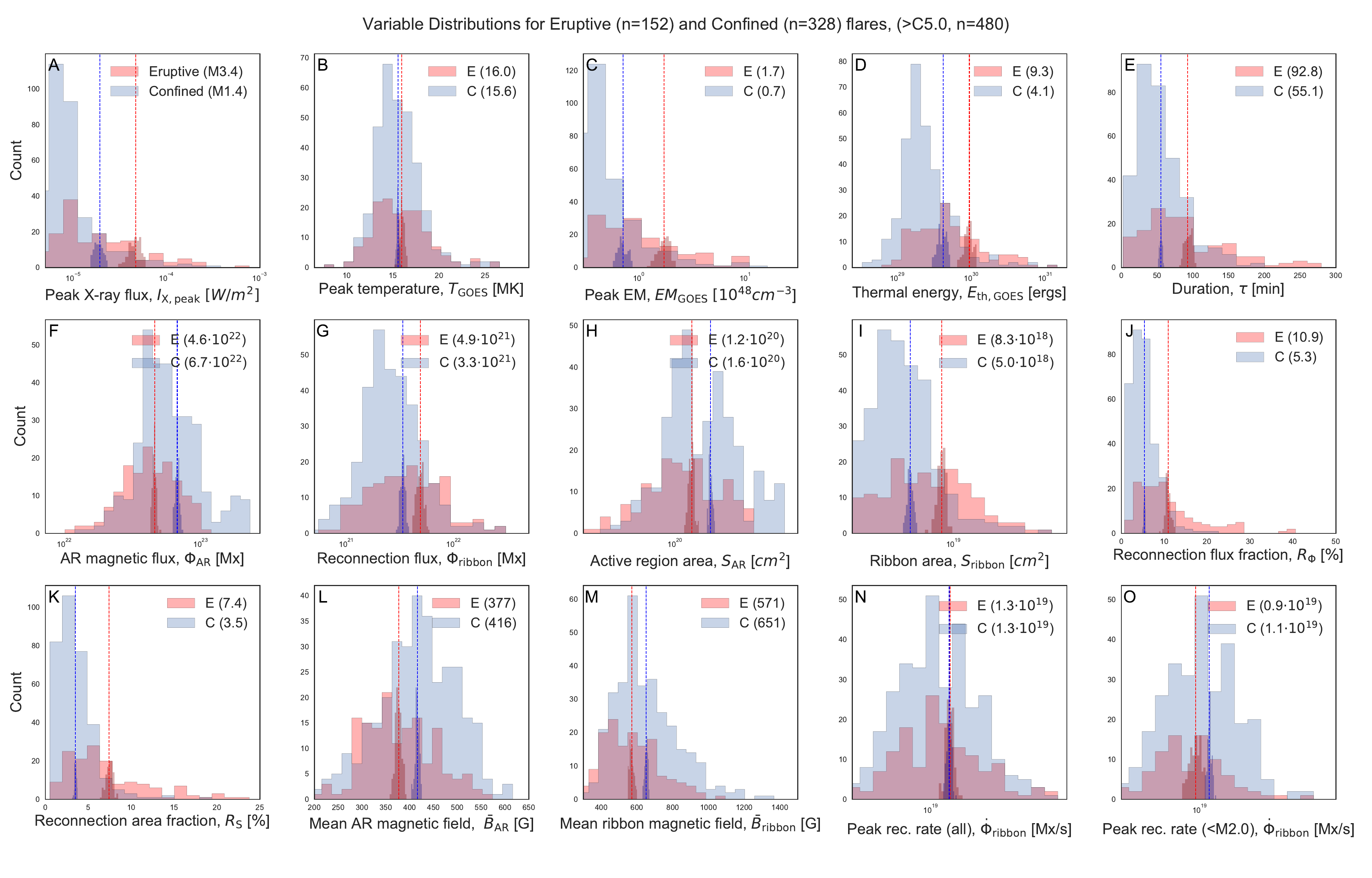}
    \caption{{Appendix: Distributions of various physical variables for eruptive (red) and confined (blue) events ($>C5.0$ GOES flare class). Vertical dotted lines and numbers in the top right corner indicate {mean value for each} variable within eruptive and confined flare groups. Darker-color histograms show distributions of the mean, if we select $100$ random subsamples of the half of the events, and reflect variability of the mean due to event selection bias. {See Appendix~\ref{app} for details.}\\}}
 \label{fig:dist_small}
 \end{figure*}
 %
\begin{figure*}[!tb]
     \includegraphics[width=0.6\textwidth]{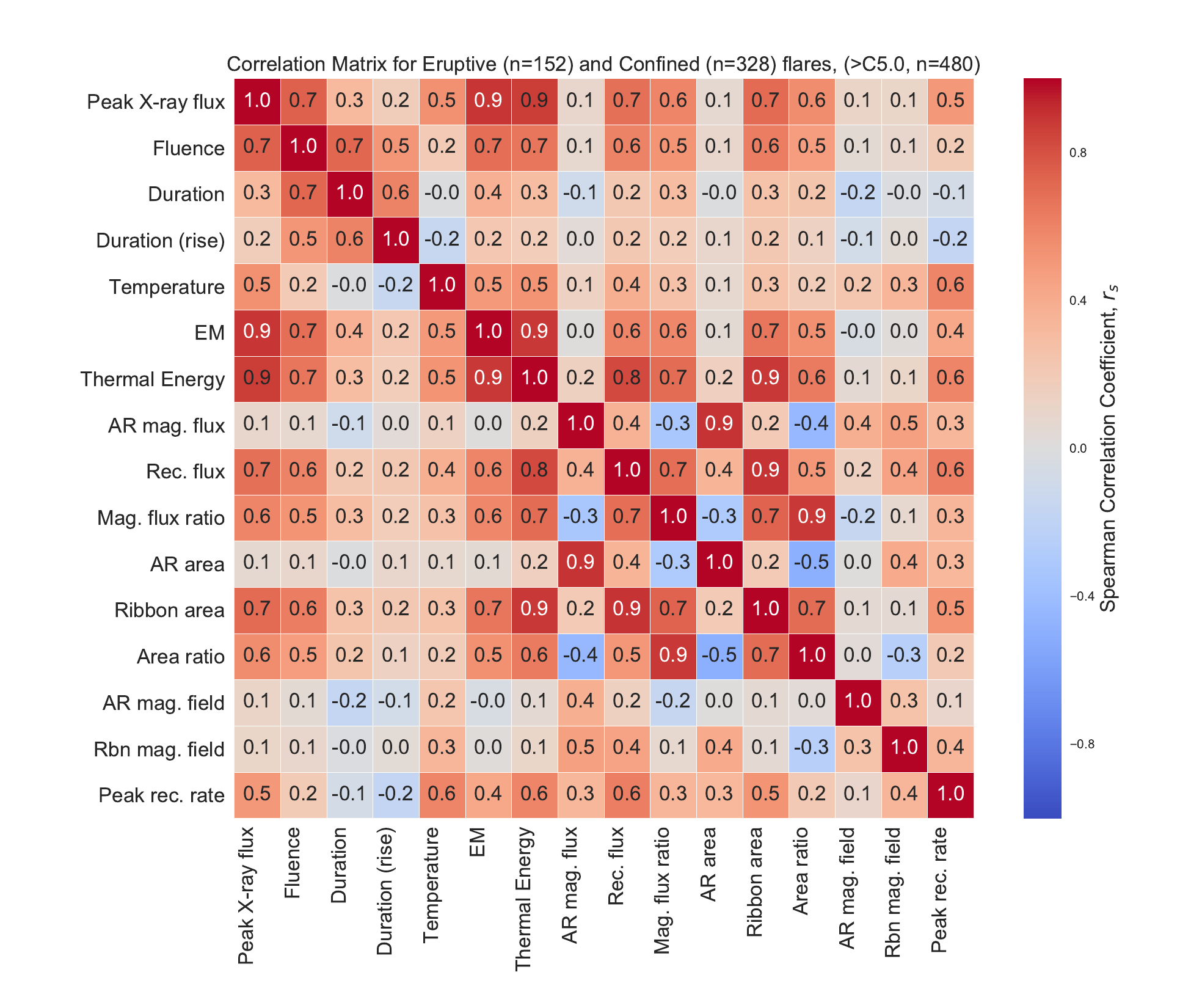}
    \caption{{Appendix: Correlation matrix showing Spearman correlation coefficient between different physical variables {for all flares with GOES flare class $>C5.0$. {See Appendix~\ref{app} for details.} }}}
 \label{fig:corrmat_small}
 \end{figure*}
\begin{figure*}[!tb]
    \includegraphics[width=1.0\textwidth]{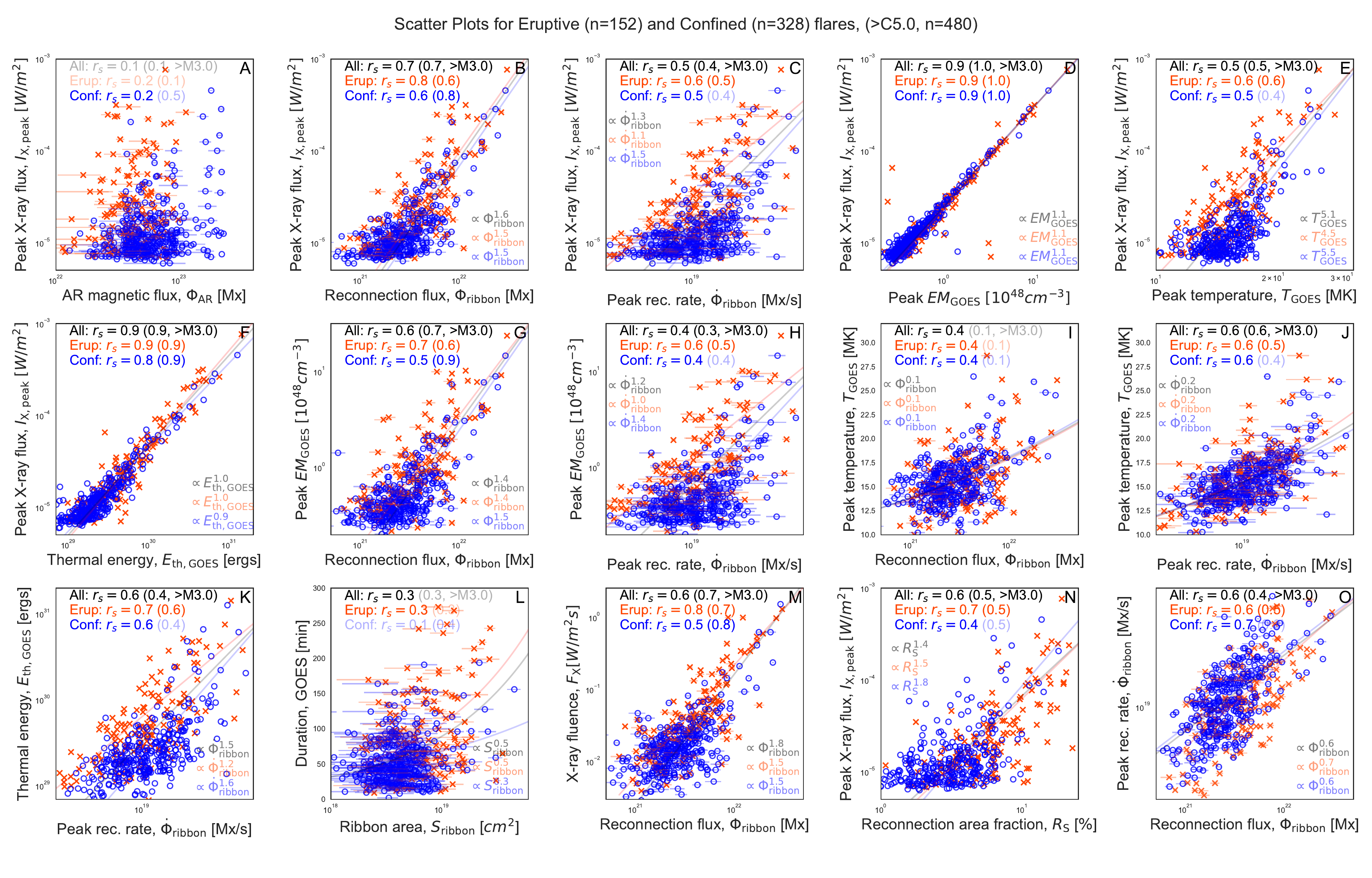}
    \caption{{Appendix: Scatter plots: between different physical variables for confined (blue) and eruptive (red) events ($>C5.0$ GOES flare class). {See Appendix~\ref{app} for details.}} \\}
 \label{fig:scat_small}
 \end{figure*}
 
In addition to flares above M1.0, as has been analyzed in the { main paper's body}, we extend our analysis to include smaller events with GOES class C5.0 and above. To minimize errors in the reconnection fluxes we restrict our analysis to events with reconnection flux imbalance between positive and negative polarities $\Phi_\mathrm{ribbon,imb}<20\%$. As a result we analyze {480 events, $152$ eruptive and $328$} confined flares, $C5.0$-class and above.

For flares of GOES class below M1.0, the fraction of confined flares significantly increases from $60\%$ for M1.0 flares to $85\%$ for C5.0 flares. As a result, the gap between distributions of GOES peak X-ray fluxes becomes even larger for confined and eruptive flares, affecting distributions of emission measure, thermal energy, etc. Nevertheless, except for mean reconnection fluxes that become smaller for confined than for eruptive flares, we observe the same tendencies between magnetic/thermodynamic properties for confined/eruptive flares as for flares above M1.0. Even the peak reconnection rates are smaller for eruptive than for confined flares, consistent with conclusions for larger flares.
 \clearpage

\end{document}